# Resonant tunneling and quantum interference of a two-spin system in silicon tunnel FETs

Satoshi Moriyama[1*], Takahiro Mori[2], Keiji Ono[3]

[1] *Department of Electrical and Electronic Engineering, Tokyo Denki University, Adachi, Tokyo 120-8551, Japan*

[2] *Device Technology Research Institute (D-Tech), National Institute of Advanced Industrial Science and Technology (AIST), Tsukuba, Ibaraki 305-8568, Japan*

[3] *Advanced Device Laboratory, RIKEN, Wako, Saitama 351-0198, Japan*

[*] E-mail: moriyama.satoshi@mail.dendai.ac.jp

We investigated the resonant tunneling of a two-spin system through the double quantum dots in Al–N-implanted silicon tunnel FETs (TFETs) by electrical-transport measurements and Landau–Zener–Stückelberg–Majorana interferometry with and without magnetic fields. Our experimental results revealed the coexistence of spin-conserving and spin-flip tunneling channels in the two-spin system in non-zero magnetic fields. Additionally, we obtained the spin-conserving/spin-flip tunneling rates of the two-spin system through the double quantum dots in the TFET. These findings will improve our understanding of the two-spin system in silicon TFET qubits and may facilitate the coherent control of quantum states through all-electric manipulation.

Quantum dots (QDs) are considered potential building blocks for future quantum devices, such as quantum bits (qubits), which operate with the coherent control of a single-electron charge and spin.[1] Thus far, we have proposed and demonstrated a high-temperature operable QD and spin-qubit based on silicon-on-insulator tunnel FETs (TFETs) with deep levels.[2] We observed room-temperature single-electron transport via the implantation-induced atomic-size QDs installed in TFETs. In previous papers,[2-4] we reported the ESR of single-electron and spin-qubit operations up to 10 K in TFET devices. Our experimental results showed that the TFET devices were operated in a Pauli spin-blockade (PSB) regime in the presence of spin-orbit coupling in double QDs.[2-5] Theoretically, for PSB under strong spin–orbit coupling,[6] estimation of the spin-flip tunneling rate from microscopic spin–orbit interaction models is the challenging issue.

In this study, we focused on resonant tunneling (RT) from the spin-singlet state to the spin-triplet states through the double QDs, which is the reverse process of the PSB regime. Here, we obtained the spin-conserving/spin-flip tunneling rates of the two-spin system in the TFET device. Furthermore, we conducted the microwave spectroscopy measurements at the RT peaks of the two-spin system. As is known, microwave spectroscopy can be employed to investigate the dynamics of the quantum states in the QD system, and the quantum coherence in the system can be evaluated by Landau–Zener–Stückelberg–Majorana (LZSM) interferometry.[7] Our analysis revealed the presence of spin-conserving and spin-flip coherent tunneling channels of the two-spin states and the coexistence of the LZSM interferences with and without magnetic fields.

All electrical transport measurements were carried out in a variable-temperature cryostat with a superconducting magnet. The measurement was performed at $T = 1.5$ K. A magnetic field of up to $B = 7$ T was applied perpendicular to the substrate, which corresponded to the [100] direction. Because the microwave amplitude delivered to the QDs depends on the cryostat setup and varies with the microwave frequency, we did not calibrate absolute power levels of the QDs. Here, the incident microwave power was labeled according to the microwave generator settings at 300 K.

Figure 1(a) shows a schematic cross-section of the TFET device with deep-level impurities in the transport channel. Figure 1(b) shows a schematic band diagram of the double QD transport system via deep-level states in the short-channel TFETs. TFETs appear similar to MOSFETs but have different types of source and drain electrodes. Our proposal for the QD devices in TFETs is to use the mid-gap deep-level states as QDs. The deep levels are intentionally introduced into the transport channel by the ion implantation of the Al–N impurity pair in our experiment.[8-11] This impurity pair generally functions as an isoelectronic trap (IET) in silicon. Theoretical calculations suggest that the nearest neighbor Al–N is favored in silicon, the ground state lies in the silicon bandgap, and a discrete state is provided.[12,13] Recently, we also reported that beryllium and the group II element, Zn, and group VI elements, S and Se, were introduced into a Si wafer to form deep levels in a TFET.[14,15] Figure 1(c) shows a charge-stability diagram, and Fig. 1(d) shows the typical current–drain bias ($I_s$–$V_d$) characteristics of an Al–N-implanted TFET with a channel length of 70 nm. We detected an RT transport that showed a sharp peak in the $I_s$–$V_d$ curve, and the peak position was independent of the gate voltage ($V_g$) range (near ($V_d$, $V_g$) = (−0.03 V, 0.25 V)) in Fig. 1(c)) in the negative-drain-bias region. In contrast, we detected a PSB transport, in which the current was suppressed in the positive-drain-bias region enclosed by the dotted line in Fig. 1(c). These results provide evidence of the series-coupled double QD transport and indicate the formation of a two-spin state. We have discussed spin manipulation in the PSB region in previous papers.[2-4] The typical ESR response in the PSB region is shown in the inset of Fig. 1(d). The $g$-factor was estimated to be 2.00 from the ESR peak position. The data similar to Figs. 1(c), (d) and their analyses are shown in Ref.[2]; however, the information is included to make this paper self-contained. The obtained $g$-factor is varied by approximately 2% by changing $V_d$ and $V_g$ in the PSB region,[2,3] due to the Stark effect.[16]

Here, we focus on the RT that was observed in the negative-drain-bias region. Let ($m$, $n$) represent the number of electrons on the left and right sides of the two QDs. Since PSB is observed when $V_d$ is positive, the charge states are (0, 1) for $V_d$ = 0 with the same $V_g$. Subsequently, as the $V_d$ is increased negatively, the Coulomb blockade is lifted and one electron enters the right QD from the right electrode, affording the singlet state ($S$(0, 2), where $|S(0,2)\rangle = (|\uparrow\downarrow\,(0,2)\rangle - |\downarrow\uparrow\,(0,2)\rangle)/\sqrt{2}$). Next, one electron tunnels from the right QD to the left QD when the energy levels of (0, 2) and (1, 1) are aligned by adjusting the $V_d$. At this time, it can be in one of a total of four states: a singlet state ($S$(1, 1)) or triplet states ($T_-$(1, 1), $T_0$(1, 1), and $T_+$(1, 1)), where $|S(1,1)\rangle = (|\uparrow\downarrow\,(1,1)\rangle - |\downarrow\uparrow\,(1,1)\rangle)/\sqrt{2}$, $|T_-(1,1)\rangle = (|\downarrow\downarrow\,(1,1)\rangle)$, $|T_0(1,1)\rangle = (|\uparrow\downarrow\,(1,1)\rangle + |\downarrow\uparrow\,(1,1)\rangle)/\sqrt{2}$, and $|T_+(1,1)\rangle = (|\uparrow\uparrow\,(1,1)\rangle)$, respectively). Thereafter, one electron tunnels from the left QD to the left electrode, returning to the (0, 1) state. And then, one electron enters the right QD from the right electrode, leading to the (0, 2) state. An illustration of the transport mechanism in Fig. 3(a) helps us understand the transport mechanism; however, note that the triplet states degenerate in a zero magnetic field. The energy difference (exchange energy, $J$) between the $S$(1, 1) and $T_0$(1, 1) states is small ($J \sim 0$); thus, they cannot be distinguished energetically, which will be explained later in the discussion of Figs. 3(b) and (c). Therefore, in the zero magnetic field, it is possible to take four (1, 1) states (one singlet and three triplet states) from the $S$(0, 2) state.

Figure 2 shows the RT peak under microwave irradiation as a function of the detuning parameter ($\Delta\varepsilon$) and the microwave power ($P$) without magnetic fields. $\Delta\varepsilon$ corresponds to the energy-level difference between the (0, 2) and (1, 1) charge states at $B$ = 0 T, and it can be tuned by adjusting the $V_d$. We estimated the conversion factor between the $V_d$ and the energy in the dot to be 0.17 meV/mV.[2] RT occurs near $\Delta\varepsilon$ = 0, when each energy level in the QDs is aligned. Figure 2(a) shows the microwave spectroscopy of the RT at a fixed microwave frequency of $f$ = 27.68 GHz. The observed peaks under microwave irradiation were due to LZSM interference and were once discussed as photon-assisted tunneling (PAT).[7, 17, 18] When the microwave power is low, satellite peaks appear, owing to PAT processes, which involve one photon emission, i.e., a satellite peak of $\Delta\varepsilon$ > 0, or one photon absorption, i.e., a satellite peak of $\Delta\varepsilon$ < 0. As the microwave power increases, more satellite peaks appear, corresponding to the emission or absorption of multiple photons. This three-dimensional plot in Fig. 2(a) shows the clear features of the PAT pattern with the emission and absorption of up to seven photons. The energies between the adjacent peaks are consistent with the single photon energy ($hf$, where $h$ is Planck's constant), and the peak height, as a function of $P$, agrees with the theoretical square Bessel function expression.[17] Figure 2(b) shows the microwave spectroscopy of the RT at the fixed $f$ = 5.61 GHz. Here, a funnel-shaped interference pattern is formed. We have observed the PAT pattern in the high frequency region. Below 10 GHz, the funnel-shaped interference pattern is observed. These results indicate that LZSM interference is observed in our device and that a coherent tunneling channel exists in the two-spin system.

In magnetic fields, the triplet states are nondegenerate owing to the Zeeman effect. Figure 3(a) shows the schematic energy diagram of an RT regime in a fixed magnetic field. Figure 3(b) shows the magnetic-field evolution of the RT peak as a function of $\Delta\varepsilon$. Above $B$ = 1 T, the RT peak for the zero magnetic field clearly splits into three peaks. The central peak ($\Delta\varepsilon \sim 0$) does not move, whereas the other two peaks shift linearly in opposite directions in the magnetic fields. These behaviors can be interpreted from the energy diagram of Fig. 3(a). That is, the peak at $\Delta\varepsilon$ < 0 ($\Delta\varepsilon$ > 0) is due to the spin-flip tunneling processes from $S$(0, 2) to $T_-$(1, 1) ($T_+$(1, 1)), where the energy levels of $S$(0, 2) and $T_-$(1, 1) ($T_+$(1, 1)) are aligned. The central peak is due to the coexistence of spin-conserving ($S$(0, 2) to $S$(1, 1)) and spin-flip ($S$(0, 2) to $T_0$(1, 1)) tunneling processes, where the energy levels of $S$(0, 2) and $S(1, 1)/T_0(1, 1)$ are aligned. The magnetic-field evolution of the energy differences between the adjacent peaks, $\Delta\varepsilon^+$ and $\Delta\varepsilon^-$, in Fig. 3(b), corresponds to the Zeeman energy ($E_z$). Figure 3(c) shows the energy differences ($\Delta\varepsilon^+$ and $\Delta\varepsilon^-$) as a function of the magnetic field. $\Delta\varepsilon^+$ and $\Delta\varepsilon^-$ are almost equal in the magnetic fields, and the slopes of the linear fitting give $g$-factors of 1.93±0.02 and 1.95±0.02 for $\Delta\varepsilon^+$ and $\Delta\varepsilon^-$, respectively. Although the obtained values are slightly smaller than those obtained in the PSB region, these results are within a reasonable range because of the $V_g$ and $V_d$ dependence of the $g$-factor.[2, 3] Regarding the central peak, the two processes, $S(0, 2)$ to $S(1, 1)$ and $S(0, 2)$ to $T_0(1, 1)$, cannot be distinguished even in a magnetic field. Assuming that the spin-conserving process is dominant in the central peak, $\Delta\varepsilon^\pm = E_z \pm J$; therefore, $\Delta\varepsilon^+ > \Delta\varepsilon^-$ is expected.[19] However, from the experimental result in Fig. 3(c), it can be observed that $\Delta\varepsilon^+$ and $\Delta\varepsilon^-$ are nearly equal. Furthermore, the central peak in the magnetic field has a shape similar to a single-component Lorentzian. These results indicate that the exchange energy between the $S(1, 1)$ and $T_0(1, 1)$ states is close to zero. Below $B = 1$ T, no splitting of the RT peak occurs, and the central peak appears to have a shape similar to single-component Lorentzian, which includes the RT processes from $S(0, 2)$ to $S(1, 1)/ T_-(1, 1)/T_0(1, 1)/ T_+(1, 1)$. Additionally, note that we observed a slight deviation for the RT peaks with and without magnetic fields from the Lorentzian shape to the right of the resonance. This deviation is a shoulder-like current step smaller than the peak height for each main peak by one order of magnitude. This deviation is attributed to the inelastic tunneling process,[20,21] which is beyond the scope of this study. Hence, heights of the RT peaks at $\Delta\varepsilon > 0$ are slightly higher than those of the RT peaks at $\Delta\varepsilon < 0$ because of the contribution of the inelastic tunneling components.

Figures 3(d) and 3(e) show the magnetic-field dependence of the peak height and the peak width, respectively, obtained from the single-Lorentzian-peak fitting of each of the three peaks. Above $B = 1$ T, the peak height in Fig. 3(d) indicates that both spin-flip-only tunneling peaks are constant at ~0.1 nA in the magnetic fields, and the peak widths of all the RT peaks are approximately the same with slight magnetic-field dependence, as shown in Fig. 3(e). The estimated FWHM of 50–80 μeV is narrower than that at the measurement temperature (the thermal energy is ~130 μeV at 1.5 K). This indicates that the electron transport is limited by the mean stay time in the double QDs, independent of the temperatures in the source and drain electrodes.[18] The peak shape of the RT due to elastic tunneling is Lorentzian theoretically[22] and can be expressed by the following formula under our measurement condition: $I(\Delta\varepsilon) = et^2\Gamma_L / [(\Delta\varepsilon/h)^2 + \Gamma_L^2/4 + t^2(2 + \Gamma_L/\Gamma_R)]$. Because the heights of both spin-flip-only RT peaks are comparable and mean stay times in the double QDs are almost comparable for all tunneling processes, assuming that the inter-dot tunneling rate of the center barrier ($t$) is dependent on the spin-conserving or spin-flip tunneling probability is reasonable. Using the height and the width of the spin-flip-only RT peak above $B = 1$ T, we determined the tunnel-coupling energies from $S(0, 2)$ to $T_-(1, 1)$ and $S(0, 2)$ to $T_+(1, 1)$ to be $ht \sim 8$ μeV, which is almost constant in the magnetic fields. Then, the contribution of the ($S(0, 2)$ to $T_0(1, 1)$) process to the central peak can be also similar to those of other spin-flip-only RT peaks (~0.1 nA). Thus, the central peak height is constant at ~0.4 nA, and the contribution of the spin-conserving ($S(0, 2)$ to $S(1, 1)$) process can be ~0.3 nA. The sum of the contributions of the processes is 0.6 nA, which is in good agreement with the height of the RT peak for the zero magnetic field. The tunnel-coupling energy for the spin-conserving process can also be estimated under the above assumptions, yielding a value of $ht \sim 12$ μeV, which is almost constant in the magnetic fields. Here, the tunnel-coupling energy from $S(0, 2)$ to $T_0(1, 1)$ is $ht \sim 8$ μeV. The deviation of all estimated parameters is approximately ±1 μeV. The occurrence of the spin-flip process is relatively frequent and as coherent as that of the spin-conserving process in the two-spin system. In addition, the spin-conserving and spin-flip processes can be described as occurring at similar rates in the zero magnetic field. The spin-flip tunneling process is attributed to the spin–orbit coupling via deep levels in Al–N-implanted TFETs, and these results suggest that the spin–orbit coupling in the QD system is strong.[2, 3]

Figure 4 shows the microwave spectroscopy of the RT peaks at several magnetic fields and microwave frequencies. We observed the LZSM interference spectra with all three peaks separated in the magnetic field, compared with the results in Fig. 2 measured at a zero magnetic field. Since the central peak indicates the coexistence of spin-conserving/spin-flip tunneling and the left and right peaks are attributed to the spin-flip-only tunneling in the two-spin system, the results of Fig. 4 indicate that all tunneling channels are coherent and coexist with approximately the same order of magnitude, with the spin-conserving tunneling slightly surpassing the spin-flip tunneling. The reason why the satellite peaks disappear in the left peak in the high magnetic field (Figs. 4(a) and (b)) is that the channels enter the Coulomb blockade region (the channels are out of the bias window) in $\Delta\varepsilon < -0.8$ meV. Although previous studies have reported the observation of RT and coherent channels with LZSM interference in the one-spin system in GaAs QDs,[23] this study is the first to observe these phenomena in the two-spin system.

In summary, we investigated the RT in a two-spin system through the double QDs in the silicon TFET. The spin-dependent RT includes one-spin-conserving process from $S(0, 2)$ to $S(1, 1)$ and three spin-flip processes from $S(0, 2)$ to $T_-(1, 1)$, $S(0, 2)$ to $T_0(1, 1)$, and $S(0, 2)$ to $T_+(1, 1)$, and all channels are coherent and coexist. We obtained the tunnel-coupling energies for the spin-flip processes to be ~8 μeV, and for the spin-conserving process to be ~12 μeV. To the best of our knowledge, as there have been no studies on electrical transport via deep impurity levels, this is the first study to estimate the spin-conserving/spin-flip tunneling rates experimentally considering the single-electron transport via deep impurity levels. The strong spin–orbit interaction in this system suggests that electron spins can be electrically manipulated, indicating that electron spins can be controlled by varying the voltage and can be coherently controlled by applying a pulsed voltage in the TFET device. These findings will improve our understanding of the interacting two-spin system of silicon TFET qubits and may enable the coherent control of quantum states by all-electric manipulation.

**Acknowledgments**

We thank Dr. Shota Iizuka of AIST for useful discussions. This work was partly supported by JST CREST Grant No. JPMJCR1871, MEXT Quantum Leap Flagship Program (Q-LEAP) Grant No. JPMXS0118069228, Japan.

**Figure Captions**

**Fig. 1.** (a) Schematic cross-section of the TFET device with deep-level impurities in the transport channel. (b) Schematic band diagram of the double QD transport via deep-level states in the short-channel TFETs. C.B., V.B., and $E_F$ denote the conduction band, valence band, and Fermi level of the electrode, respectively. (c) Intensity mapping plot of the differential conductance ($dI_s/dV_d$) in the double QD transport regime as a function of $V_d$ and $V_g$. (d) Log-scale $I_s$–$V_d$ characteristics in the double QD transport regime. The data correspond to the cross-section of Fig. 1(c) at $V_g = 0.255$ V. Inset: ESR response under the PSB conditions at ($V_d$, $V_g$) = (0.085 V, 0.253 V). Steady-state change in $\Delta I_s$ as a function of the magnetic fields with $f = 16.55$ GHz and $P = 3$ dBm.

**Fig. 2.** Three-dimensional plot of $I_s$ as a function of the detuning parameter ($\Delta\varepsilon$) and incident microwave power ($P$). Here, $P$ was labeled with microwave generator settings at 300 K. The microwave frequencies were fixed at (a) 27.68 and (b) 5.61 GHz. The left inset figure shows $I_s$ as a function of $\Delta\varepsilon$ at a fixed $P$, corresponding to the cross-section of each main figure. For clarity in the inset of Fig. 2(a), each trace is shifted in −0.15 nA steps from the bottom to the top.

**Fig. 3.** (a) Schematic energy diagram of a resonant tunneling regime of two-spin states at a fixed magnetic field. $\Gamma_L$, $t$, and $\Gamma_R$ represent the tunneling rate of each barrier. (b) $I_s$ as a function of $\Delta\varepsilon$ in a magnetic field. Each trace is shifted in 0.2 nA steps from the bottom to the top for clarity. (c) Magnetic-field dependence of the energy differences ($\Delta\varepsilon^+$ (upper triangle) and $\Delta\varepsilon^-$ (under triangle)). The solid and dashed lines are the linear fitting for $\Delta\varepsilon^+$ and $\Delta\varepsilon^-$, respectively. (d) Magnetic-field dependence of the peak height of three RT peaks. The upper triangles are those from $S(0, 2)$ to $T_-(1, 1)$, open circles are those from $S(0, 2)$ to $S(1, 1)/T_0(1, 1)$, and open squares are those from $S(0, 2)$ to $T_+(1, 1)$. The closed circles for $B = 0$ T and 0.5 T are extracted from the central-peak data because the peak does not split. (e) Magnetic-field dependence of FWHM of the three RT peaks. The assignment of symbols is the same as in (d).

**Fig. 4.** Intensity mapping plot of $I_s$ as a function of $\Delta\varepsilon$ at (a) $B = 7.0$ T and $f = 32.70$ GHz,

(b) $B = 7.0$ T and $f = 16.96$ GHz, (c) $B = 3.0$ T and $f = 32.70$ GHz, and (d) $B = 1.0$ T and $f = 16.96$ GHz. To clearly observe the PAT pattern of each peak, the energy relationship $E_z > hf$ was selected for the measurement.

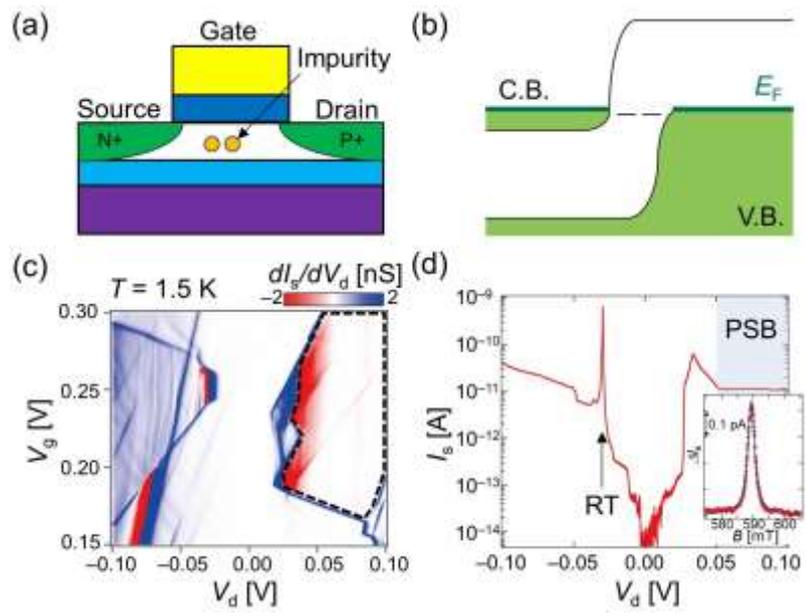

**Fig. 1.**

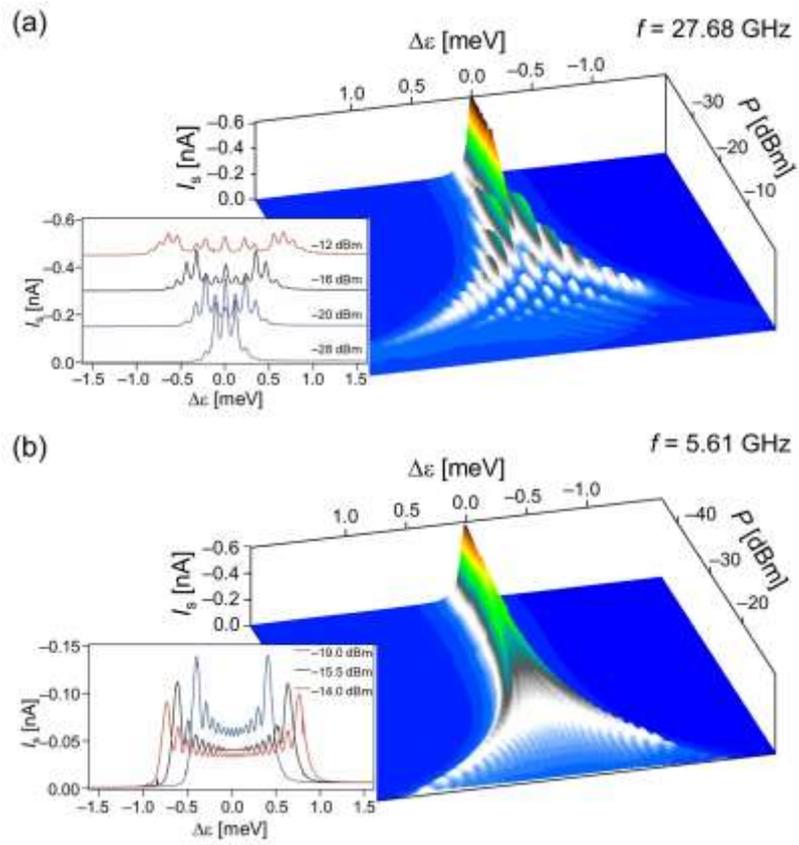

**Fig. 2.**

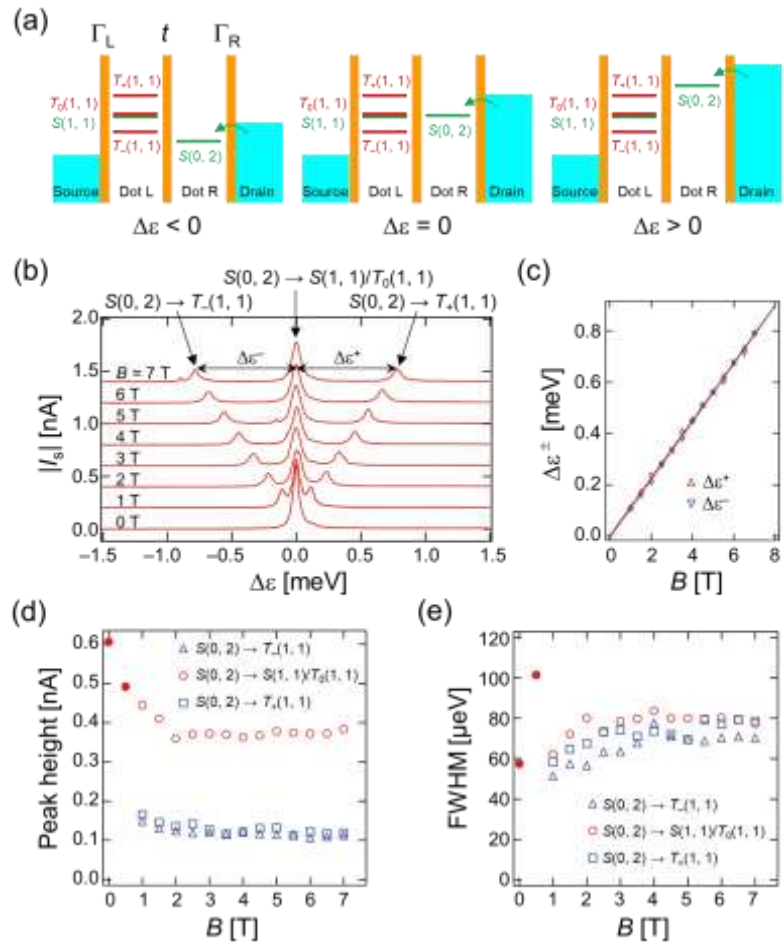

**Fig. 3.**

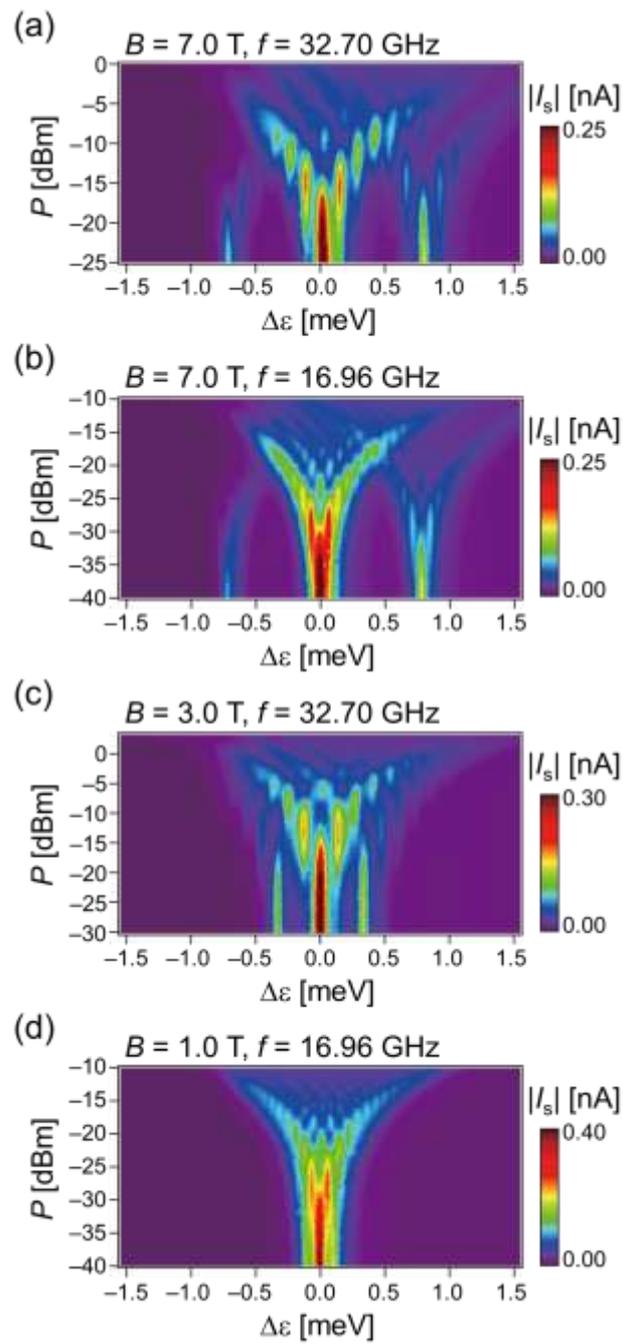

**Fig. 4.**